\documentclass[sigconf,nonacm]{acmart}

\AtBeginDocument{%
  }

\newif\ifcomm
\commfalse
\ifcomm
\newcommand{\EA}[1]{\textcolor{green}{EA:~#1}} 
\newcommand{\SK}[1]{\textcolor{purple}{SK:~#1}} 
\newcommand{\AP}[1]{\textcolor{violet}{AP:~#1}} 
\else
\newcommand{\EA}[1]{}
\newcommand{\SK}[1]{}
\newcommand{\AP}[1]{}
\fi

\begin{document}

\title[Towards Automatic Hands-on-Keyboard Attack Detection Using LLMs in EDR Solutions]{Towards Automatic Hands-on-Keyboard Attack Detection Using LLMs in Endpoint Detection and Remediation (EDR) Solutions}

\author{Amit Portnoy}
\authornote{Both authors contributed equally to this research.}
\email{amitportnoy@microsoft.com}
\orcid{0000-0001-6491-5814}
\author{Ehud Azikri}
\authornotemark[1]
\email{ehudazikri@microsoft.com}
\author{Shay Kels}
\email{shkels@microsoft.com}
\affiliation{
  \institution{Microsoft Security AI Research}
}









\begin{abstract}
Endpoint Detection and Remediation (EDR) platforms are essential for identifying and responding to cyber threats. This study presents a novel approach using Large Language Models (LLMs) to detect Hands-on-Keyboard (HOK) cyberattacks. Our method involves converting endpoint activity data into narrative forms that LLMs can analyze to distinguish between normal operations and potential HOK attacks. We address the challenges of interpreting endpoint data by segmenting narratives into windows and employing a dual training strategy. The results demonstrate that LLM-based models have the potential to outperform traditional machine learning methods, offering a promising direction for enhancing EDR capabilities and apply LLMs in cybersecurity.
\end{abstract}



\received{28 May 2024}

\maketitle

\section{Introduction}

Cybersecurity has become an ever-present concern in our increasingly digital world. As organizations and individuals become more reliant on technology, the impact of cyberattacks has grown in both frequency and severity. In recent years, attackers have often focused on \textit{Hands-on-Keyboard} (HOK) attacks, where adversaries engage directly with their target's systems. These attacks are often meticulously planned and executed, allowing attackers to maneuver through networks undetected, making them a considerable challenge for security professionals \cite{liveaction2022, crowdstrike2023, microsoftlearn2022}.  
  
Traditionally, cybersecurity detection methods focused on matching known patterns of malicious activity against observed events. However, this approach is less effective against HOK attacks, which frequently employ novel techniques and can vary significantly between instances. Anomaly detection systems offer an alternative by identifying deviations from baseline behaviors, yet they too struggle with the subtlety and sophistication of HOK attacks, often resulting in high false positive rates.  
  
The limitations of these conventional methods have spurred interest in more advanced and adaptive security measures. In recent years, the rise of machine learning (ML) and artificial intelligence (AI) has provided new tools for threat detection and response. ML-based systems can learn from data to identify complex patterns and predict future events, offering the potential to recognize and react to HOK attacks more effectively \cite{buczak2015survey, sahani2023machine}.  
  
Among the various AI technologies, \textit{Large Language Models} (LLMs) have emerged as a particularly promising avenue for cybersecurity applications. These models, exemplified by OpenAI's GPT series \cite{10.5555/3495724.3495883}, have demonstrated remarkable abilities in natural language understanding and generation. By training on vast corpora of text, LLMs can infer context, identify relationships between entities, and even emulate human reasoning to some extent \cite{liu2023evaluating}. The prospect of applying LLMs to cybersecurity hinges on their potential to process and interpret the vast amounts of unstructured text data generated by security systems, such as logs, alerts, and reports. Having a deeper understanding and stored knowledge, LLMs can identify narratives and patterns that are indicative of HOK activity, which might be overlooked by traditional systems.  
  
In this paper, we explore the integration of LLMs into \textit{Endpoint Detection and Response} (EDR) solutions to enhance the detection of HOK cyberattacks. We propose a novel approach that transforms endpoint data into structured narratives, which we term "endpoint stories." These stories are designed to capture the essence of security events, providing a coherent and contextualized account that can be analyzed by LLMs to discern between benign operations and potential threats.  
  
The complexity and volume of endpoint data, the need for models that can interpret technical and domain-specific language, and the requirement to maintain high accuracy while minimizing false positives all present significant hurdles. Furthermore, the operational deployment of LLMs and their response latency must be carefully considered, given the computational resources required and the critical need for real-time analysis in cybersecurity operations.  

This work addresses these challenges and demonstrates the viability of using LLMs for HOK attack detection. The integration of LLMs into cybersecurity practices represents not just a technical enhancement but a paradigm shift in how we conceptualize and implement security measures. Our main contributions include: (a) A new way to transform endpoint behavior into human-readable text admissible to processing by methods leveraging modern LLMs; and (b) An LLM based pipeline for classification of text extracted from endpoint behaviors that provides a practical solution for dealing with long context-lengths through a separate training of embeddings and the classification head.

In Section \ref{sec:related_work}, we provide a review of related work, highlighting past approaches to cyberattack detection and the evolution of AI and LLMs in cybersecurity. Section \ref{sec:methodology} details our methodology, encompassing data collection, preparation, and the development of our LLM-based detection framework. Specifically, in Sub-section \ref{subsec:data}, we delve into the specifics of our data handling practices, ensuring both the richness of the data for model training and adherence to privacy standards. Section \ref{sec:evaluation} describes the experimental setup, including the model training process, the evaluation metrics employed, and the testing protocols. It includes an analysis of the performance of our approach in detecting HOK attacks and compares it to traditional methods. Finally, section \ref{sec:conclusion}, concludes the paper by summarizing our contributions and envisaging the impact of our research on the future landscape of cybersecurity defenses.

\section{Related Work}  \label{sec:related_work}
  
The domain of cybersecurity has witnessed a surge in research efforts aimed at detecting and mitigating Hands-on-Keyboard (HOK) cyberattacks. Previous studies have emphasized the importance of detecting these threats due to their targeted and interactive nature, which often results in significant damage to affected networks and systems \cite{bdcc7030143, razaulla2023age}. Traditional approaches to HOK attack detection have focused on signature-based methods, which rely on known patterns of malicious activity, and anomaly detection techniques, which aim to identify deviations from normal behavior \cite{jeffrey2023review}.  
  
With the increasing sophistication of cyberattacks, these conventional methods have faced challenges in keeping pace with the evolving tactics employed by attackers. This has led to the exploration of machine learning and artificial intelligence (AI) as tools for enhancing detection capabilities. In particular, supervised and unsupervised machine learning models have been employed for pattern recognition and to learn from historical data, enabling more proactive and dynamic defense mechanisms \cite{bdcc7030143}.  
  
Recently, the application of Large Language Models (LLMs) in cybersecurity has opened new avenues for research. LLMs, such as OpenAI's GPT-3 \cite{10.5555/3495724.3495883}, have demonstrated remarkable capabilities in understanding and generating human-like text, which can be leveraged to interpret the contextual information within security logs \cite{boffa2024logprecis, sai2024generative}. LLMs have also been instrumental in automating threat intelligence tasks, such as phishing email identification and malware classification based on textual descriptions \cite{hafzullah2024llm, hassanin2024comprehensive}. These advancements suggest that the integration of LLMs into cybersecurity tools could significantly enhance the ability to detect complex threats such as HOK attacks.  

Despite the potential of recent LLMs, in relation to HOK detection or related intrusion detection, current work focus on custom models that lack wide knowledge and are relatively tiny in number of parameters and context length \cite{10393509, 10423646}. This is mainly due to the computational resources required to train and deploy LLMs at the scale in which such systems operate.

Our work is the first to our knowledge that attempts tapping into the full potential of modern LLMs and present a novel methodology that not only harnesses their power for detecting HOK attacks but also addresses the practical considerations for deployment within an endpoint protection platform. In the next section, we present our model architectures and efficient training techniques that maintain high performance while reducing resource demands.

\section{Methodology}  \label{sec:methodology}
  
Our approach to detecting Hands-on-Keyboard (HOK) cyberattacks employs a novel integration of Large Language Models (LLMs) with endpoint security data. This section outlines the comprehensive methodology we adopted, encompassing data preparation, collection, and the application of LLMs for early attack detection.  
  
\subsection{Data Preparation and Collection}  \label{subsec:data}
  
The foundation of our methodology lies in the careful preparation and collection of endpoint security data. We began by accumulating a vast dataset of endpoint logs, which consist of detailed records of the following types of events:

\begin{itemize}
    \item \textbf{Telemetry}---raw events, such as process creation, file operations (create/delete/move), registry operations, logon attempts, etc.
   \item \textbf{Security observations}---Manual expertise based security events that aggregate some telemetry events and represent low fidelity signals of possible malicious activity but are not enough to raise an alert without further evidence. These observations correlate with many that exist in MITRE ATT\&CK framework \cite{mitre_attack}, but are highly curated for our purpose. An example of such an observation could be multiple failed login attempts from the same user during a brief time span.
   \item \textbf{Machine learning (ML) observations}---Scores of proprietary ML models aimed to detect types of malicious activities obtained during that time frame. These include suspicious command lines, registry activity, etc.
\end{itemize}

The data has undergone a rigorous anonymization process that strips away any personally identifiable information (PII) while preserving the essential characteristics of the events for analysis.  

Once the data was anonymized, we embarked on the process of transforming these logs into a more structured and consumable format, which we refer to as "endpoint stories." These narratives are designed to concisely convey the sequence and context of events, facilitating the subsequent application of LLMs. The transformation process involves several key steps: 

\begin{figure*}[t]
   \centering
   \includegraphics[width=\textwidth]{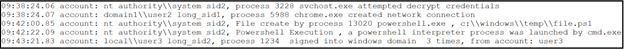}
   \caption{An example of an endpoint story segment. \SK{Could we improve the readability of the example?}}
   \label{fig:endpoint_story}
\end{figure*}

\begin{itemize}
    \item \textbf{Evidence Aggregation}---aggregate the evidence from the endpoint during a timeframe and sort events by the time of occurrence.
\item \textbf{Filtering}---To reduce the story length, we filter events that, based on expert knowledge, do not contribute to HOK classification. 
\item \textbf{Rephrasing}---rephrase the evidence to a consistent format starting with user information, followed by evidence type and source and terminated by specific evidence details.
\item \textbf{Deduplication}---Deduplicating the evidence by temporal/source logic. To further reduce the endpoint story size, similar evidence (e.g., all the process creation events triggered by the same parent process) occurring in a single second are grouped and formatted as single evidence.
\item \textbf{Normalization}---Normalizing evidence content by enumeration. To preserve consistency and reduce story size, references of long entities, such as file names, process, etc.

\end{itemize}

For training a classification model, we collect malicious and benign training examples. In view of the immense throughput of our EDR solution, the endpoint stories are generated when there is an initial and low-confidence indication of a possible HOK attacks, such as a single EDR alert. The malicious training examples correspond to endpoint stories taken from machines that correspond to HOK incidents that have been reviewed and validated by security experts. Benign training examples are collected from endpoints for which there was only one EDR alert in the preceding days, and this alert was not indicative of an HOK attack. Benign examples are not manually reviewed due to volume, thus may have a labeling noise to some small degree.

To obtain results indicative of the true performance of the model, we split the data into training and test by time. The training set contains $\sim$8K examples and is equally balanced. The test set contains $\sim$2.7K equally balanced examples. Figure \ref{fig:endpoint_story} includes an example of a few lines taken from an endpoint story.

\subsection{Model Architecture and Training}  

With the endpoint stories prepared, we proceeded to the model training phase. We utilized a state-of-the-art LLM that was further fine-tuned on our specific dataset to ensure high relevance to the cybersecurity domain. The model was trained to classify endpoint stories into benign or potential HOK attack categories. 

To train our models over long endpoint stories we split our text into windows, we use a pre-trained LLM to create a per-window embedding and then pass all the windows through an additional classification head. We evaluated two approaches for the training process: (1) training the window model and the classification head in tandem; and (2) training a model for creating per-window embedding and separately training a model for classifying a sequence of window embeddings.

When training the windows with the classification head, we use the well-known BERT-small model \cite{devlin-etal-2019-bert}, which is smaller and easily fits in GPU memory with an entire security log sample served as a single batch of $\sim$80 windows. For the classification head, we attach a bidirectional LSTM layer of each window CLS token.

For the second approach, we fine-tune the Phi-2 model \cite{Javaheripi2023}, which is a general purpose LLM that is still small enough to process multiple windows with low latency (2.7B parameters). Since Phi-2 is a generative model, we suffix each window with a few tokens designed to extract expressive embedding. Specifically, we append \texttt{‘[[classification:negative’} or  \texttt{‘[[classification:positive’} for each window depending on whether it originates from a clean (negative) or an HoK attack (positive) sample. We then take the embedding of the colon (\texttt{‘:’}) token from the last layer as a representation of the window. For training the classification head, we concatenate the embeddings of all windows and pass them as a sequence to a BERT-like model with 4 transformer layers with 10 attention heads in each. For comparison purposes, we also describe a classifier that only uses the fine-tuned Phi-2 base model and outputs the average probability of predicting the \texttt{‘positive’} token over all the windows of the story.
  
  
  

\section{EVALUATION}  \label{sec:evaluation}
  
We evaluate our models over data collected from global production EDR deployments labeled as described in the previous sections. Our primary metric for assessing a model for HOK attacks detection is True-Positive Rate (TPR, aka, recall) at False-Positive Rate (FPR) of 1\%. We enforce a low FPR because EDR operates in an extremely high volume and false positives translate to a very high cost and friction with clients. For completeness, we also include the \textit{Area Under the Receiver Operating Characteristic Curve} (AUC) score, we emphasis it is not appropriate for our business needs as it does not take into consideration our focus on precision over recall.

As a baseline classification model, we compare our models to LIghtGBM, which is a well-known Gradient boosting technique, which does not account for the token order in the text. We also test two additional techniques to gain insight on our approach. In the first, "Phi-2 + Avg. \texttt{'positive'} token", we run our fine-tuned Phi-2 over each models and take the score of predicting \texttt{'positive'} as the next token. We then take the average of those scores. With this techniques we effectively only learn over windows of endpoint stories. In the second technique we just takes the average score of both our BERT-bases and Phi-2-based models. 

\begin{table}[h]
\centering
\begin{tabular}{|l|c|c|}
\hline
\textbf{Model} & \textbf{AUC} & \textbf{TPR at FPR 1\%} \\ \hline
LightGBM & 0.977 & 0.524 \\ 
BERT + LSTM head & 0.968 & 0.694 \\
Phi-2 + Avg. \texttt{'positive'} token & 0.929 & 0.463 \\
Phi-2 + Transformer-head & 0.971 & 0.725 \\
Average of BERT + Phi-2 & \textbf{0.979} & \textbf{0.754} \\ \hline
\end{tabular}
\caption{Model Performance Metrics}
\label{table:model_performance}
\end{table}

The results of the study presented in Table \ref{table:model_performance}. We note that the "Phi-2 + Avg. \texttt{'positive'} token" method produced reasonable results but it is considerably inferior to other methods which suggest that many windows simply do not have enough data for accurate prediction. The LightGBM model, while having a high AUC, has a lower TPR at FPR 1\% compared to the other models, indicating that it may not be as effective in detecting HOK attacks at high precision. Phi-2 with the transformer is notably the best standalone model followed by the BERT + LSTM head model, while both considerably outperform LightGBM in our low-FPR regime. Lastly, BERT + Phi-2 average has the highest AUC and TPR at FPR 1\%. This demonstrates that even a simple combination of  LLMs can lead to improved performance and suggests that ensemble methods should be considered whenever possible.

\section{Conclusion}  \label{sec:conclusion}
  
Our research represents a critical step forward in the detection of Hands-on-Keyboard (HOK) cyberattacks. The use of Large Language Models (LLMs) to analyze structured narratives, or "endpoint stories," has proven to be a promising approach in distinguishing between benign and malicious activities within an endpoint protection framework. Our experimental results have demonstrated that LLMs can outperform traditional detection methods, providing a higher degree of accuracy while maintaining a manageable false positive rate.  
  
Throughout this study, we have addressed several challenges, including the transformation of complex endpoint data into a format amenable to LLM processing, the development of an effective training pipeline which allow for long context input. Our research has shown that it is possible to leverage the power of advanced machine learning techniques to improve cybersecurity defenses significantly.  
  
Looking ahead, there are numerous avenues for future work. Enhancing the interpretability of LLM decisions, reducing the resource requirements for model training and inference, and further refining the models to adapt to the evolving tactics of cyber adversaries are all critical areas for continued research and experimenting with alternative techniques for long-context detection.

\bibliographystyle{ACM-Reference-Format}
\bibliography{sample-base}

\appendix









\end{document}